# MP-SeizNet: A Multi-Path CNN Bi-LSTM Network for Seizure-Type Classification Using EEG.

Hezam Albaqami[a,b,∗], Ghulam Mubashar Hassan[a] and Amitava Datta[a]

[a]*Department of Computer Science and Software Engineering, The University of Western Australia, Australia*
[b]*Department of Computer Science and Artificial Intelligence, University of Jeddah, Saudi Arabia*

ABSTRACT

Seizure type identification is essential for the treatment and management of epileptic patients. However, it is a difficult process known to be time consuming and labor intensive. Automated diagnosis systems, with the advancement of machine learning algorithms, have the potential to accelerate the classification process, alert patients, and support physicians in making quick and accurate decisions. In this paper, we present a novel multi-path seizure-type classification deep learning network (MP-SeizNet), consisting of a convolutional neural network (CNN) and a bidirectional long short-term memory neural network (Bi-LSTM) with an attention mechanism. The objective of this study was to classify specific types of seizures, including complex partial, simple partial, absence, tonic, and tonic-clonic seizures, using only electroencephalogram (EEG) data. The EEG data is fed to our proposed model in two different representations. The CNN was fed with wavelet-based features extracted from the EEG signals, while the Bi-LSTM was fed with raw EEG signals to let our MP-SeizNet jointly learns from different representations of seizure data for more accurate information learning. The proposed MP-SeizNet was evaluated using the largest available EEG epilepsy database, the Temple University Hospital EEG Seizure Corpus, TUSZ v1.5.2. We evaluated our proposed model across different patient data using three-fold cross-validation and across seizure data using five-fold cross-validation, achieving F1-scores of 87.6% and 98.1%, respectively.

## 1. Introduction

Chronic diseases, including epilepsy, are a growing concern to human health [1]. Epilepsy is a neurological disorder characterized by abnormal and abrupt neuronal discharges in the brain (seizures) [2]. After stroke, it is the most common brain disorder in humans [3]. Seizures discharges weaken the brain's normal functioning, exposing the patient to dangerous situations. Severe seizures limit a person's ability to engage in independent and social activities, leading to repercussions such as social isolation and low educational and work achievements [4]. The most severe effects of epileptic seizures include an increased risk of worst outcomes like death [4, 5].

There are many types of epileptic seizures, and each type necessitates a different therapy approach [6]. Anti-epileptic drugs (AEDs), such as primidone and phenytoin, are usually used to control seizures [7]. However, they might not be effective in drug-resistant epileptic patients, for whom surgical intervention would be required [8]. Moreover, it has been stated that for one-third of epileptic patients, no medical treatment options exist [6, 9]. These patients need to determine ways to live with their disease and organize their everyday lives in a manner that takes their condition into account.

The accurate and early identification of epileptic seizures serves as the foundation for all critical diagnostic and therapy procedures and is the starting point of the treatment process for epilepsy. To ensure that treatment processes are efficacious, epileptic seizures must first be correctly identified because some medications are only suitable for the intended specific condition [10].

The most typical approach for diagnosing epileptic seizures is a spontaneous electroencephalogram (EEG) with electrodes attached to the scalp; however, it is associated with certain difficulties [11]. In addition to being a labor-intensive and time-consuming process, deciphering EEG data requires a highly qualified neurologist. According to [12], the ratio of neurologists to patients is quite low, especially in low-income countries which may lead to a significant delay in diagnosis. Even in developed countries, it has been reported that patients with certain types of seizures experience a long delay in being notified about their prognosis, which in turn causes worse seizure outcome [13].

Furthermore, the clinical and EEG characteristics of different types of seizures are in many respects similar to one another, which may lead to misinterpretations regarding the seizure type, making diagnosis more difficult. Correctly identifying the exact type of seizure through EEG data constitutes a significant challenge, even for a highly trained neurologist [14]. They are required to incorporate patient or carer diaries into their diagnoses, but these diaries are always subject to mistakes and forgetfulness [15].

Additionally, video-EEG monitoring (VEM) is required to support the diagnosis processes when a diagnosis cannot accurately be made on clinical grounds [11]. VEM is a technique that concurrently records EEG data, video, and sounds to capture a patient's behavior during seizures. This process entails patients being admitted to a monitoring room for a long period to record spontaneous or provoked seizure events. Furthermore, it calls for further manual analysis. Therefore, there is a dire need to develop an automated method for classifying epileptic seizure types to facilitate and expedite arduous

∗Corresponding author:
haalbaqamii@uj.edu.sa ( Hezam Albaqami)

diagnostic procedures and enhance patient care quality.

In this study, a novel architecture for effectively classifying seizure types using EEG signals is proposed. A multi-path seizure-type classification deep learning network (MP-SeizNet) was trained on different representations of EEG data for epileptic seizure type classification. The proposed network is based on a convolutional neural network (CNN) and a bi-directional long short-term memory (Bi-LSTM). The latter is used to extract complex patterns from the raw EEG signals. The CNN is used to learn from a handcrafted set of features computed using different wavelet-based feature extraction methods, namely wavelet packet decomposition (WPD), dual-tree complex wavelet decomposition (DTCWT), and discrete wavelet transform (DWT). Our MP-SeizNet fuses the outputs of both CNN and Bi-LSTM for the final classification of seizure types.

## 1.1. Review of related work

Works on automated seizure type classification have predominantly focused on either feature-based or end-to-end approaches [16]. Feature-based methods depend on careful and manual feature engineering. They have a long history of successful application in a wide range of EEG analyses, such as in brain–computer interface (BCI) technology, including motor imaginary task classification [17] and spatial attention shifts detection [18] and also for the diagnostic evaluation of various brain-related disorders (e.g., seizure detection) [3, 19].

The strategies used in EEG feature extraction include time domain [20, 21], frequency domain [6, 22], and time-frequency domain features [18, 23, 24]. Time-frequency methods, such as wavelet transform and short-time Fourier transform (STFT)-based analysis, are commonly used in EEG analysis. Recently, a combination of wavelet transform and statistical approaches for EEG feature extraction has gained popularity among researchers [3, 17, 18, 25, 26, 27, 28, 29].

On the other hand, end-to-end methods accept raw data with minimal prepossessing fed into the deep learning algorithms to extract features from the input data automatically. This approach was inspired by the high level of performance of applications in other research fields, such as computer vision and speech recognition. For example, for EEG analysis, Schirrmeister et al. [30] proposed a CNN model to decode imagined and executed tasks from multi-channel raw EEG data, achieving competitive results to classical feature-based solutions. Similarly, Altan et al. [31] introduced a novel method using deep learning with CNN to predict imagined left and right movements through raw EEG data. The results show that the proposed prediction model outperforms conventional fully-connected neural networks and achieves high classification performances.

For diagnostic evaluation of various brain-related conditions using EEGs, the authors in [32] proposed a recurrent neural network (RNN) architecture, ChronoNet, to detect abnormal EEG recordings using the representation of raw EEG data. In addition, a CNN model showed its strength to discriminate between pathological and normal subjects using raw EEG data [33].

Another popular practice in utilizing deep learning methods for EEG analysis is by representing the input signal as a different representation to be fed into the deep learning model. Spectrogram image representations of signals based on STFT, fast Fourier transform (FFT), and continuous wavelet transform (CWT) have also been found valuable by many researchers. Additionally, autoencoder (AE) models are popular unsupervised deep learning methods that generate different input data representations, which are then classified using different supervised learning methods. Altan et al. [34] proposed a deep learning framework that generates a different presentation of the data using AE and Extreme Learning Machine (ELM) for the classification of brain activity successfully.

Deep learning methods using multi-representation of data have recently gained popularity, especially for cross-subject analysis [35]. This technique employs more than one data representation to a model for more comprehensive learning [36]. Then, in a multi-path scenario, the output of each path is integrated into a single and compact output using different methods, such as point-wise addition and concatenation [37]. It has been stated that deep learning models with multi-representation of data learn more comprehensive features than individual representation [37, 38]. Therefore, in this paper, we employ two different representations of the seizure EEG data, namely time-frequency domain features and the raw EEG signals, for the problem of seizure type classification.

For seizure type classification, there are different methods used for feature extraction and classification using the aforementioned techniques. Table 1 presents recent works for EEG-based epileptic seizure types classification.

Sriram et al. [39] classified the EEG signals into eight different seizure types using a popular deep learning model AlexNet with transfer learning. Similarly, Raghu. S et al. [40] applied the method of transfer learning on images extracted from EEG signals to classify seizures using the InceptionV3 model, and the study reported an accuracy of 88.3%. Liu et al. [41] proposed a novel hybrid bilinear model consisting of a CNN and LSTM to classify spectrogram images extracted from EEGs for seizure type identification, achieving an F1-score of 97.4%. In addition, a Plastic Neural Memory Network (NMN) was proposed for seizure classification [42]. The study reported an F1-score of 94.5%. Shankar et al. [43] classified the EEG signals into five classes of seizures using a novel technique based on Gramian Angular Field Transformation (GAFT) and CNN. The study reported accuracies of 96.01%, 89.91%, 84.19% and 84.20% for 2,3,4 and 5 class classification problems, respectively. However, the study utilized an oversampling technique that can be unfavorable for real-world scenarios.

Baghdadi et al. [44] proposed a novel channel-wise attention-based deep LSTM model for identifying eight types of seizures, achieving an F1-score of 98.41%. Similarly, Priyasad et al. [45] employed an attention-driven data fusion technique for seizure classification from raw EEG data, reporting an

**Table 1**
Summary of existing literature on automated multi-class seizure type classification.

| Study | Year | Method | Class | Patient-dependent Results | Patient-dependent Validation | Patient-independent Results | Patient-independent Validation |
|---|---|---|---|---|---|---|---|
| Sriram et al. [39] | 2019 | • AlexNet, <br>• Transfer learning approach, <br>• Short-time Fourier transforms (STFT), <br>• Spectrogram images. | 8 classes* | 84.06 Acc | Hold-out testset | — | — |
| Saputro et al. [46] | 2019 | • Support Vector Machine (SVM) <br>• Mel Frequency Cepstral Coefficients (MFCC), <br>• Hjorth Descriptor, <br>• Independent Component Analysis (ICA). | 4 classes* | 91.4 Acc | Hold-out testset | — | — |
| Wijayanto et al. [47] | 2019 | • SVM, <br>• Empirical Mode Decomposition (EMD), <br>• Intrinsic Mode Functions (IMFs), <br>• Statistical Features. | 5 classes | 95 Acc | Hold-out testset | — | — |
| Saric et al. [25] | 2020 | • Multi-layer perceptron (MLP), <br>• Continuous Wavelet Transform (CWT), <br>• Statistical Features. | 3 classes* | Hold-out testset | 95.14 Acc | — | — |
| Roy et al. [6] | 2020 | • XGBoost. <br>• Eigenvalues of Fast Fourier Transform (FFT) <br>• Correlation matrix *IBMFS*. | 7 classes | 90.1 F1 | 5-fold CV | 54.2 | 3-fold CV |
| Raghu. S, et al. [40] | 2020 | • Inception v3, <br>• Transfer learning. <br>• STFT spectrogram Images. | 8 classes* | Hold-out testset | 88.3 Acc | — | — |
| Liu et al. [41] | 2020 | • Hybrid bilinear CNN + Long Short-Term Memory (LSTM), <br>• STFT spectrogram images, | 8 classes | 97.4 F1 | 5-fold CV | — | — |
| Asif et al. [48] | 2020 | • Ensemble Learning SeizureNet CNNs, <br>• FFT spectrograms. | 7 classes | 95 F1 | 5-fold CV | 62 | 3-fold CV |
| Aristizabal et al. [42] | 2020 | • Plastic Neural Memory Network (NMN), <br>• IBMFS [6]. | 7 classes | 94.5 F1 | 5-fold CV | — | — |
| Shankar et al. [43] | 2021 | • CNN, <br>• Gramian Angular Field Transformation (GAFT). | 5 classes* | 84 Acc | Hold-out testset | — | — |
| Albaqami et al. [26] | 2021 | • LightGBM, <br>• Wavelet Packet Decomposition (WPD), <br>• Statistical features. | 7 classes | 89.6 F1 | 5-fold CV | 64 | 3-fold CV |
| Naze et al. [49] | 2021 | • Random Forests. <br>• Spectrum Power Bands, <br>• Principal Component Analysis (PCA), | 2 classes | 3-fold CV | 94.3 Acc | — | — |
| McCallan et al. [27] | 2021 | • Bagged trees classification. <br>• Discrete Wavelet Decomposition (DWT), <br>• Statistical features. | 6 classes | 82 Acc | Hold-out testset | — | — |
| Jia et al. [50] | 2022 | • ResNet18_v2 with variable weight CNN algorithm (VW̄CNN), <br>• IBMFS [6] | 7 classes | 94 F1 | 5-fold CV | 54.2 Acc | Hold-out testset contains 16 patients |
| Zhang. S et al. [51] | 2022 | • Non-linear twin SVMs (NLTWSVM), <br>• Variational Mode Decomposition (VMD), <br>• IMFs Statistical features. | 7 classes | 92.3 F1 | 5-fold CV | — | — |
| Albaqami et al. [28] | 2022 | • LightGBM, <br>• Dual Tree Complex Wavelet Transform (DTCWT), <br>• Statistical features. | 5 classes | 99.1 F1 | 5-fold CV | 74.7 | 3-fold CV |

*Including normal EEG label.

**Table 1** (continued)

| Study | Year | Method | Class | Performance Evaluation(%) | | | |
|---|---|---|---|---|---|---|---|
| | | | | Patient-dependent | | Patient-independent | |
| | | | | Results | Validation | Results | Validation |
| Baghdadi et al, [44] | 2022 | • LSTM<br>• Channel-wise attention mechanism,<br>• Raw EEGs. | 8 classes | 98.41 F1 | 5-fold CV | — | — |
| Priyasad et al.,[45] | 2022 | • Channel-wise shallow SincNet+CNN<br>• Attentive fusion mechanism<br>• Raw EEGs. | 8 classes | 96.7 F1 | 5-fold CV | – | – |
| Tang et al., [52] | 2022 | • Recurrent Graph Neural Network (GNN),<br>• Geometry and connectivity of EEG electrodes,<br>• STFT. | 4 classes | – | – | 74.9 F1 | Hold-out testset. |
| Shankar et al., [53] | 2022 | • CNN+LSTM,<br>• Hilbert vibration decomposition (HVD)<br>• Continues Wavelet Transform (CWT), | 6 classes* | 98.8 F1 | Hold-out testset | — | — |

*Including normal EEG label.

F1-score of 96.7%. The technique involves using independent shallow networks, SincNet, trained on each EEG channel. The output from each network is then fused using an attention mechanism to extract the salient information for classification. However, the channel-wise attention mechanism in [44, 45] is computationally expensive, especially if the number of channels is large. Indeed, a common limitation of previous research studies is that they have yet to experiment with applying their techniques across different subjects, which is crucial for evaluating generalization ability.

Asif et al. [48] proposed a novel deep learning framework named at SeizureNet, consisting of stacked CNNs for ensemble learning. The study reported a high performance of 95% F1-score at the seizure level. However, the performance of SeizureNet dramatically decreased when evaluated across different patients and reduced to 62% F1-score. In [50] a variable weight CNN algorithm (VWCNN) with ResNet was utilized to classify features extracted from EEGs into seven classes of seizures. The study reported an F1-score of 94% across seizures and an accuracy of 54.2% when their model generalized to data from unseen patients. As stated before, Table 1 presents the current literature on the problem of seizure type classification.

### 1.2. Motivation and contributions

Based on our literature review, the published research on seizure type classification has two major shortcomings. First, the patient-specific situation has been the primary focus of the current and previous research, while the patient-non-specific scenario has rarely been considered. Patient-specific scenarios tackle the problem of decoding pathology from EEGs based on a patient's own historical data. In the patient-non-specific scenario, the learning methods are applied across different patients. The former always achieves a high level of accuracy since the training and testing samples are derived from the same source and have comparable distributions. On the contrary, when these methods are generalized to data from new sources, specifically new patients, they reveal glaring shortcomings. Based on these aforementioned limitations, we thus present in this work a robust technique that can learn common seizure patterns from different patient data towards greater generalizability and reliability in diagnoses. We rigorously examined our proposed technique as applied to different subjects in addition to standard testing to guarantee its effectiveness.

Second, a very large portion of previous research that proposed solutions to automate the process of seizure type classification did not explicitly follow the nomenclature of seizure types. According to the International League Against Epilepsy (ILAE), epileptic seizures in the updated ILAE 2017 guidelines can be classified into focal onset, generalized onset, and unknown onset [54]. Each type can be extended to more specific seizure types based on electrographic and clinical symptoms [55]. A common available seizure dataset, Temple Hospital University (TUH) EEG Seizure Corpus TUSZ, includes a variety of EEG seizure recordings labeled based on the 1981 ILAE seizure classifications [55, 56]. The recordings in the dataset are classified into simple partial seizure (SPZ), complex partial seizure (CPZ), absence seizure (ABZ), tonic seizure (TNZ), tonic-clonic seizure (TCZ), and myoclonic seizure (MYZ). If there was no clear evidence to label a recording into one of the specific seizure types, the recording was classified either as focal non-specific or generalized non-specific [56]. In the literature, some studies, if not most, have tried to classify the entire dataset into one of the eight classes, regardless of the medical terminology associated with those labels. Consequently, this may diminish the quality of the obtained results. Therefore, in this paper, we present an automated framework for seizure type classification with consideration of the medical terminology regarding the types of seizures in the commonly used dataset.

To date, as per the best of our knowledge, previous research on the problem of seizure type classification has only tackled the problem from one side, either using a feature-based approach or an end-to-end approach in one single representation (see Table 1). In this paper, we combine both approaches as two different representations of EEG data for seizure type classification with a novel technique that cap-tures discriminatory features from raw EEG signals and handcrafted features. Our proposed network, MP-SeizNet, harnesses the power of wavelet-based features and deep learning algorithms to automatically discover important features for the task of seizure type classification.

The following are the key novel contributions of this study:

- We propose a novel seizure type classification deep learning network, MP-SeizNet, that takes advantage of time-frequency EEG wavelet-based features and signals' temporal complex patterns using a multi-path neural network.

- Our proposed solution incorporates DWT, DTCWT, WPD and CNN to obtain salient EEG features that can discriminate seizure types effectively. Similarly, our proposed network utilizes both past and future signal information by employing Bi-LSTM with an attention mechanism for more comprehensive feature learning.

- We evaluate our proposed model across different patients using the world's largest EEG dataset, TUSZ v1.5.2., containing various seizure morphologies.

- The proposed MP-SeizNet achieved the best results in the patient-non-specific scenario and we attribute this to the importance of integrating both the wavelet-based features and the raw EEG data in the learning process.

- To the best of our knowledge, this is the first study to incorporate both handcrafted features and raw EEG data as two different representations for seizure type classification using a multi-path deep learning network.

The remaining parts of this article are organized in the following fashion. Section 2 explains the methodology of our proposed solution, including the dataset, preprocessing steps, and a brief overview of proposed network's components. Section 3 describes the evaluation methods and the conducted experiments, the parameter settings, and the obtained results. Section 4 discusses, in general, the results of the current work and its limitation. Finally, Section 5 concludes this article with future work recommendations.

## 2. Methodology

Our study proposes to use the advancement in machine learning algorithms for the problem of seizure types classification using EEG. This section explains the detailed implementation steps of our novel proposed MP-SeizNet.

### 2.1. EEG Data

The TUSZ v1.5.2 [56] serves as the basis for our study since it is the largest publicly available multi-class seizure EEG dataset. It was collected from real-world hospital data at Temple University Hospital (TUH) between 2002 and 2012. The TUSZ is currently the only dataset of its kind, including various seizure labels recorded from over 300 unique patients and more than 3000 seizure events. The seizure events in the TUSZ dataset are labeled with the following labels:

- Focal Non-specific Seizure (FNZ),
- Generalized Non-specific Seizure (GNZ),
- Simple Partial Seizure (SPZ),
- Complex Partial Seizure (CPZ),
- Absence Seizure (ABZ),
- Tonic Seizure (TNZ),
- Tonic-clonic Seizure (TCZ),
- Myoclonic Seizure (MYZ).

Table 2 presents the dataset labels' definitions and statistics in detail. In this study, we exclusively focus on the problem of the specific seizure type classification because the non-specific seizure labels found in the dataset are not entirely disjoint from the specific labels, as stated in previous research studies [42, 26, 28, 52]. In addition, we exclude seizure type MYZ since there are only three such cases recorded from two patients in the TUSZ v1.5.2 (see Table 2). This decision is also in line with the earlier studies focusing on seizure type classification [6, 40, 48, 42, 26, 28, 52].

### 2.2. Proposed technique

This subsection explains the pre-processing steps, feature extraction, and proposed network. The proposed network also introduces brief overview of CNN, Bi-LSTM and attention mechanism.

#### 2.2.1. Preprocessing

The scalp EEG recordings in the TUSZ were collected at different sampling rates, ranging between 250 and 512 Hz, and the data was recorded based on the International 10-20 system electrode placement map [57]. Therefore, we followed the prior studies' preliminary preprocessing measures to normalize the input distribution, assure data consistency, and provide stability in the network learning and feature extraction process [6, 48, 26]. The preprocessing are summarized as follows:

1. Use the transverse central parietal (TCP) montage, including 20 EEG channels. Fig. 1 illustrates the channels used in this study.
2. Re-sample all recordings at 250 Hz if it is not.
3. Crop each recording into equally non-overlapped segments such that each segment is of the length of two seconds.

#### 2.2.2. Feature extraction

Before the feature extraction and after the initial preprocessing steps (see Section 2.2.1), the EEG signals were passed through a bandpass filter with a cut-off frequency of 0.1–50 Hz, which was inspired by [26]. After that, three different techniques of wavelet decomposing: DWT, DTCWT, and WPD are applied to the EEG signals from which different statistical features are extracted from each decomposed signal. We computed six statistical features that have been successful in the literature in discriminating the EEG signals [26, 23, 24, 29]. The computed features are:

**Table 2**
TUSZ's statistics, labels and their medical definitions.

| | Seizure Type | Definition | ILAE 2017 | No. of seizure events | Duration (Seconds) | No. of patients |
|---|---|---|---|---|---|---|
| **Focal Seizure** | FSZ | Focal seizure event that lacks information to determine its specific type. | Focal seizures | 1836 | 121139 | 150 |
| | CSZ | Focal seizure event during unconsciousness which is identified by clinical symptoms only. | Focal Impaired Awareness | 367 | 36321 | 41 |
| | SPZ | Focal seizure event during consciousness which is identified by clinical symptoms only. | Focal aware | 52 | 2146 | 3 |
| **Generalized Seizure** | GSZ | Generalized seizure event that lacks information to determine its specific type | Generalised Seizures | 583 | 59717 | 81 |
| | ASZ | Absence discharges observed on EEG, the patient experiences brief periods of unconsciousness. | Absence | 99 | 852 | 12 |
| | TNZ | Sudden stiffening of muscles during seizures. | | 62 | 1204 | 3 |
| | TCZ | Sudden stiffening and then jerking or twitching of muscles | Tonic-Clonic | 48 | 5548 | 12 |
| | MYZ | Brief seizures with muscles and limbs jerking. | Myoclonic | 3 | 1312 | 2 |

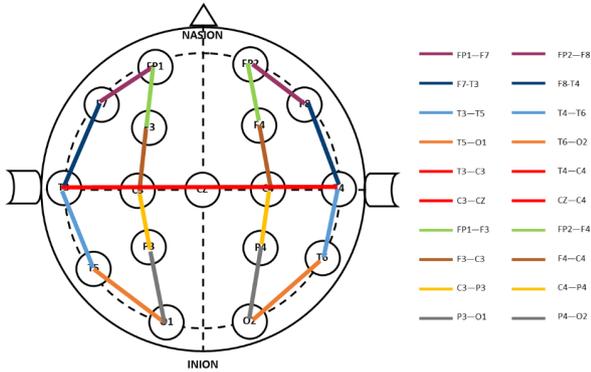

**Fig. 1**: TCP montage-based 20 EEG channels used in this study, along with their respective locations on the scalp [26].

1. Mean absolute values $F1$,
2. Average power $F2$,
3. Standard deviation $F3$,
4. Ratio of the absolute mean values of adjacent coefficients $F4$,
5. Skewness $F5$,
6. Kurtosis $F6$.

Table 3 presents each technique's computational complexity, the decomposition levels, the number of the selected coefficients and the number of extracted features. The choice of the number of the decomposition level and the number of features are based on successful research studies on the same topic [26, 28].

**Table 3**
Number of features in each decomposition technique

| Method | Computational complexity | No. of levels | No. of sub-bnds | No. of features |
|---|---|---|---|---|
| DWT | $O(N)$ [58] | 5 | 6 | 30 |
| DTCWT | $O(N)$ [59] | 5 | 6 | 30 |
| WPD | $O(N \log N)$ [58] | 5 | 32 | 192 |
| | Total number of extracted features: | | | 252 |

We computed a feature vector $F$ of 252 (see Table 3) per channel. The resultant feature vector dimension is $N$ x $C$ x $F$ x 1, where $N$ is the number of EEG samples, $C$ is the number of EEG channels and $F$ is the number of computed features.

### 2.3. Computational complexity of features

The computational complexity of our features is presented in this section using the big-O notation ($O$) method [60]. The big O notation gives an upper bound on the growth rate of the running time of an algorithm as the input size increases. It provides the researchers with enough information to identify the most appropriate feature set to use [61]. The actual complexity could be affected by different factors such as specific implementation, the size of the dataset and the underlying framework.

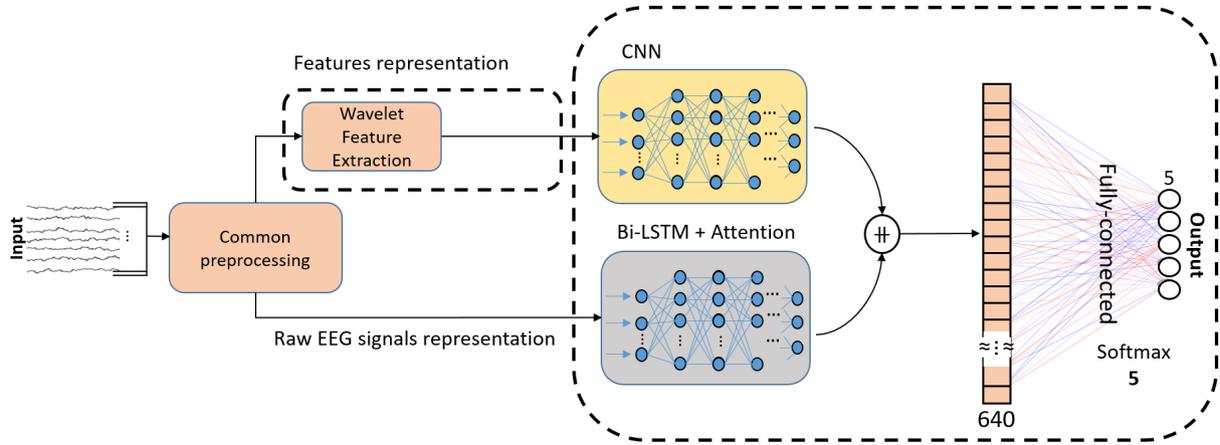

**Fig. 2**: Block diagram of our proposed MP-SeizNet for the problem of classifying the specific type of seizure. The MP-SeizNet consists of two sub-models: CNN sub-model which is trained on time-freqency representation of the signals using wavelet-based features, and Bi-LSTM sub-model which is trained on raw time-series EEG data. The outputs of each sub-models are concatenated and passed through a fully-connected layer with a softmax activation function for classification.

Table 3 presents the computational complexity of each wavelet decomposition technique utilized in our research. The computational complexity of the computed statistical features is $O(N)$ for each feature [62]. Therefore the total computational complexity of all processes will be $O(N) + O(N) + O(N \log N) + O(N) \approx O(N \log N)$.

### 2.4. Proposed deep learning network

Our proposed MP-SeizNet employs different representations of the signals to learn the hidden patterns of the seizure type within the signals for more robust performance and generalization capability. In this study, two different representations of the EEG data are used, and two deep learning sub-models are built for each type of input, CNN and Bi-LSTM sub-models. The first representation is the wavelet-based features (see Section 2.2.2), which are given as input to the first sub-model, the CNN. The second representation is the raw EEG signals with the initial preprocessing (see Section 2.2.1), fed to the second sub-model, the Bi-LSTM. Our proposed MP-SeizNet combines the output of each sub-model and is trained entirely on the two different representations of the signals. Fig. 2 shows the block diagram of or proposed model. In the following subsections, we present the detailed steps of both sub-models and the proposed MP-SeizNet.

#### 2.4.1. CNN sub-model

Convolutional neural networks (CNNs) have achieved remarkable performance in recent years on various challenging problems such as computer vision, speech recognition and natural language processing (NLP). For example, CNNs have been able to do some difficult visual tasks better than humans [63]. Furthermore, they are not restricted to specific types of input data but have emerged with promising performance in various fields, for example, but not limited to signal processing in brain-computer interface (BCI) and biomedical signal processing [30, 64]. The main idea of CNNs is that they can learn a set of hierarchical features that comprises high-level features out of low-level features through convolution [63].

The proposed CNN sub-model consists of two blocks of convolutions, followed by a max-pooling layer and a fully-connected layer, as shown in Fig. 3. Each block contains four convolutional layers and one max-pooling layer. The outputs of the first two convolutional layers are concatenated, and provided as an input to a third convolutional layer followed by a max-pooling layer, which is further followed up with a final convolution layer. Spatial Dropout and Batch normalization were used between and after the convolutional layers. The choices of the kernel, kernel size, and the number of convolutional layers were chosen empirically based on trial and error and inspired by the choices in [64].

In summary, our final CNN sub-model comprises eight convolutional layers, three 2-dimensional max-pooling layers, and one fully connected layer. In addition, our sub-model uses Batch Normalization (BN), Concatenation, Dropout, and Spatial-Dropout to speed up the learning process, preserve the learned features, and avoid over-fitting. Fig. 3 and Table 4 present the details of the architecture of our CNN sub-model.

As discussed in Section 2.2.2, the wavelet-based extracted features are given as input to the CNN sub-model to learn the hidden complex pattern for seizures. Since the deep learning models are data-hungry and require lots of data samples, we passed the extracted data without any feature elimination or selection.

#### 2.4.2. Bi-LSTM sub-model

LSTMs are specifically implemented to maintain long-term data sequences with a memory cell structure. Memory cell, also know as cell state, is the core idea behind the LSTM where cell states have the ability to control how the information in a sequence of data enters, stored, and leaves the network with gates activation functions. Three gates namely input, forget and output gates are contained in the memory cell as presented in Fig. 4. LSTM cell's state is spilt into two vectors short-term and long-term states, $h_{(t)}$ and $c_{(t)}$ respectively. The long-term state from previous time step $c_{(t-1)}$, travel through the network from left to right passing first through the forget gate to drop some memories, controlled by $f_{(t)}$, and then adding new memories that were selected via the input gate that is controlled by the gate controller $i_{(t)}$. The input gate receives its input $g_{(t)}$ which is the analyzed current input $x_{(t)}$ and the previous short-term state $h_{(t-1)}$. After that, the resultant $c_{(t)}$ is sent out without any further computation. Also, the long-term state $c_{(t)}$ after the addition operation is copied and passed through the *tanh* function which is then filtered by the output gate, controlled by the gate controller $o_{(t)}$, that produces the new short-term state $h_{(t)}$ as well as the current cell's output which is $y_{(t)}$. The $f_{(t)}$, $g_{(t)}$ and $o_{(t)}$ are three gate controllers using logistic activation function producing values ranging between 0 to 1 to open or close the gates [63].

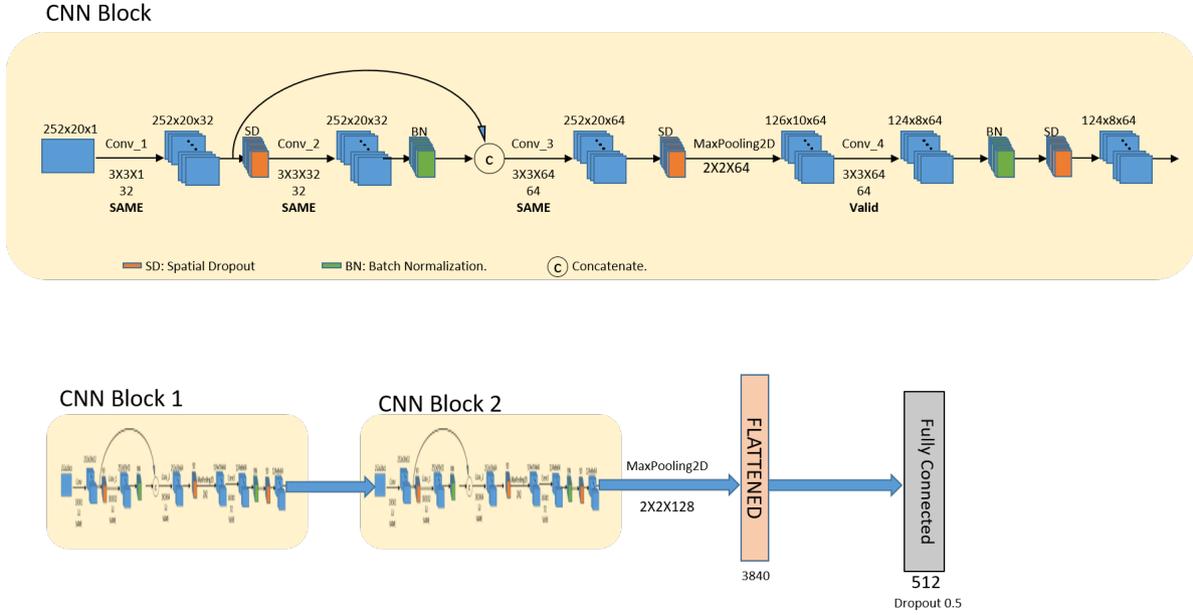

Fig. 3: Block diagram of the proposed CNN sub-model. The model consists of eight convolutional layers, three 2D-Max-pooling, two fully connected layers, six spatial dropout layers and a regular dropout.

**Table 4**
Implementation details for our proposed CNNs model.

| | Layer (type) | Maps | Size | Kernel Size | Stride | Padding | No. Parameters |
|---|---|---|---|---|---|---|---|
| | Input (Input) | 1 | 252×20 | — | — | — | 0 |
| | Conv_1 (Conv2d) | 32 | 252×20 | 3×3×1 | 1 | SAME | 320 |
| | Activation (LeakyReLU) | 32 | 252×20 | – | – | – | 0 |
| | SD_1 (SpatialDropout2D) | 32 | 252×20 | – | – | – | – |
| | Conv_2 (Conv2d) | 32 | 252×20 | 3×3×32 | 1 | SAME | 9248 |
| | BN_1 (BatchNormalization) | 32 | 252×20 | – | – | – | 128 |
| CNN BLOCK 1 | Activation (LeakyReLU) | 32 | 252×20 | – | – | – | – |
| | Concatenation | 64 | 252×20 | – | – | – | – |
| | Conv_3 (Conv2d) | 64 | 252×20 | 3×3×64 | 1 | SAME | 36928 |
| | SD_2 (SpatialDropout2D) | 64 | 252×20 | – | – | – | – |
| | Maxpooling2d | 64 | 126×10 | 2×2X64 | 2 | – | – |
| | Conv_4 (Conv2d) | 64 | 124×8 | 3×3×64 | 1 | VALID | 36928 |
| | BN_2 (BatchNormalization) | 64 | 124×8 | – | – | – | 256 |
| | Activation (LeakyReLU) | 64 | 124×8 | – | – | – | – |
| | SD_3 (SpatialDropout2D) | 64 | 124×8 | – | – | – | – |
| | Conv_5 (Conv2d) | 64 | 124×8 | 3×3×64 | 1 | SAME | 3692 |
| | Activation (LeakyReLU) | 64 | 124×8 | – | – | – | – |
| | SD_4 (SpatialDropout2D) | 64 | 124×8 | – | – | – | – |
| | Conv_6 (Conv2d) | 64 | 124×8 | 3×3×64 | 1 | SAME | 36928 |
| | BN_3 (BatchNormalization) | 64 | 124×8 | – | – | – | 256 |
| CNN BLOCK 2 | Activation (LeakyReLU) | 64 | 124×8 | – | – | – | – |
| | Concatenate_1 | 128 | 124×8 | – | – | – | – |
| | Conv_7 (Conv2d) | 128 | 124×8 | 3×3×128 | 1 | SAME | 147584 |
| | SD_5 (SpatialDropout2D) | 128 | 124×8 | – | – | – | – |
| | Maxpooling2d | 128 | 62×4 | 2×2X128 | 2 | – | – |
| | Conv_8 (Conv2d) | 128 | 60×2 | 3×3×64 | 1 | VALID | 147584 |
| | BN_4 (BatchNormalization) | 128 | 60×2 | – | – | – | 512 |
| | Activation (LeakyReLU) | 128 | 60×2 | – | – | – | – |
| | SD_6 (SpatialDropout2D) | 128 | 60X2 | – | – | – | – |
| | Maxpooling2d | 128 | 30×1 | 2×2×128 | 2 | – | – |
| Classification | Flatten | – | 3840 | – | – | – | – |
| | Dense | – | 512 | – | – | – | 1966592 |
| | Dropout | – | – | – | – | – | – |
| | Class output (softmax) | – | 5 | – | – | – | 565 |

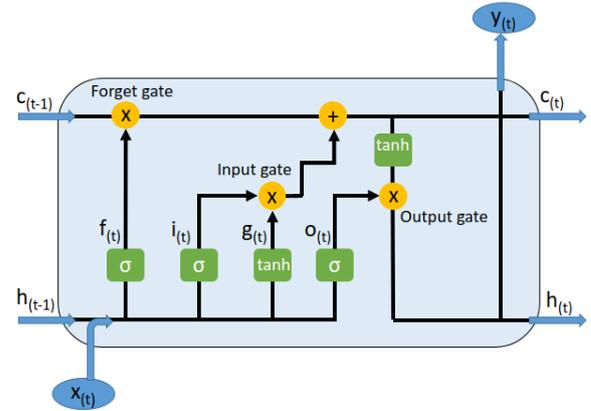

Fig. 4: Internal structure of the LSTM cell containing three gates: Forget, Input and output

Therefore, the network learns what memories to keep, what to forget and what to read from it. The following equations summarize the computation of the cell's long-term state $c_t$, its short-term state $h_t$ and its output $y_t$ at each time step $t$:

$$\mathbf{i}_{(t)} = \sigma\left(\mathbf{W}_i \cdot \left[\mathbf{x}_{(t)}, \mathbf{h}_{(t-1)}\right] + \mathbf{b}_i\right) \quad (1)$$

$$\mathbf{f}_{(t)} = \sigma\left(\mathbf{W}_f \cdot \left[\mathbf{x}_{(t)}, \mathbf{h}_{(t-1)}\right] + \mathbf{b}_f\right) \quad (2)$$

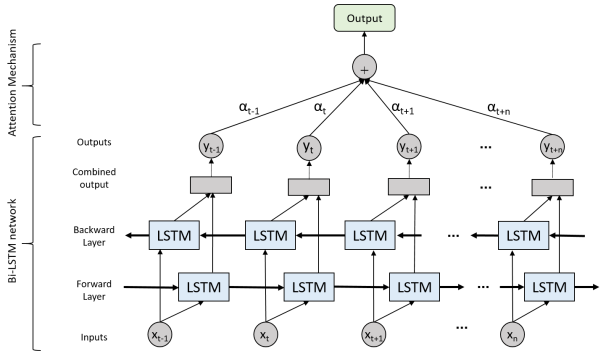

**Fig. 5**: Schematic diagram of proposed Bi-LSTM with attention mechanism, showing the output of the forward and backward layer at each stage is given to activation layers that form the relation of past and future information and how this output passed through attention mechanism to get the most influential features.

$$\mathbf{o}_{(t)} = \sigma \left( \mathbf{W}_0 \cdot [\mathbf{x}_{(t)}, \mathbf{h}_{(t-1)}] + \mathbf{b}_o \right) \tag{3}$$

$$\mathbf{g}_{(t)} = \tanh \left( \mathbf{W}_g \cdot [\mathbf{x}_{(t)}, \mathbf{h}_{(t-1)}] + \mathbf{b}_g \right) \tag{4}$$

$$\mathbf{c}_{(t)} = \mathbf{f}_{(t)} \otimes \mathbf{c}_{(t-1)} + \mathbf{i}_{(t)} \otimes \mathbf{g}_{(t)} \tag{5}$$

$$\mathbf{y}_{(t)} = \mathbf{h}_{(t)} = \mathbf{o}_{(t)} \otimes \tanh \left( \mathbf{c}_{(t)} \right) \tag{6}$$

where $W_i, W_f, W_o, W_g$ are the weight matrices, and $\mathbf{b}_i \mathbf{b}_f, \mathbf{b}_o$ and $\mathbf{b}_g$ are the bias terms.

LSTMs only process the data sequence in one direction which is utilising the past information. Bidirectional LSTM (Bi-LSMT) networks are extensions of conventional LSTM in which they process the input data in two directions: froward as well as backward. This type of network can be considered as two LSTMs, one processing the sequence of data from past to future and the other processing the data from future to past as depicted in Fig. 5. The output of each LSTM at the current time step $t$, $\vec{h}_t$ and $\overleftarrow{h}_t$ are combined to form the output of the Bi-LSTM:

$$h_t = \vec{h}_t \mathbin{+\mkern-10mu+} \overleftarrow{y}_t \tag{7}$$

where $\mathbin{+\mkern-10mu+}$ denotes the concatenation mode used to merge the output of each LSTM layers to form one output, therefore doubling the output size of Bi-LSTM.

In this study, our second sub-model is a Bi-LSTM sub-model that processes EEG data directly without any prior feature extraction process. The raw EEG data is fed directly to the Bi-LSTM in the shape of (Timeseries x EEG channel). We are utilizing 20 EEG channels (see Section 2.2.1), and each sample is a duration of two seconds sampled at 250 Hz. Our Bi-LSTM sub-model consists of one layer of Bi-LSTM with number of neuron set to 64. An attention layer followed the Bi-lstm layer is to get the most important features from the Bi-LSTM as discussed in detail below.

### 2.4.3. Attention Mechanism

Attention mechanism is commonly used in various research fields such as image captioning, NLP, and signal processing [65]. The attention mechanism place a thorough focus on some of the most important information, while the redundant and irrelevant information is dropped away. The idea of the attention mechanism is based on weight distribution, allocating higher weights to the most useful and effective information. It can improve the performance of Bi-LSTM by focusing on certain time-steps with the most discriminatory information [65]. Therefore, in this study instead of directly taking the output of the last hidden state $h_t$ of the Bi-LSTM, the output is passed through an attention layer that place a through focus on the most distinguishing features by multiplying the output $h_t$ by trainable parameters as shown in Fig. 5. In three phases, the attention weights are computed as follows:

$$s_t = \tanh \left( W_h h_t + b_h \right) \tag{8}$$

$$\alpha_t = \frac{\exp(s_t)}{\sum_t \exp(s_t)} \tag{9}$$

$$s = \sum_t \alpha_t h_t \tag{10}$$

where $h_t$ is the input to the attention layer. $s_t$ is attention score, $W_h$ and $b_h$ are the trainable parameters weight and bias of the attention layer, respectively. The obtained score $s_t$ is passed through activation softmax function to ensure that all these weights add up to 1. Finally, the attention adjusted output state $s$ is calculated by weighted sum as presented in Eq. (10).

We used an attention layer after the Bi-LSTM to get the most salient features at each time step.

### 2.4.4. Proposed MP-SeizNet

Fig. 2 presents the complete architecture of the proposed MP-SeizNet for the problem of seizure type classification. The outputs of each sub-models are concatenated and given as input to the final softmax classification layer. It is important to note that the learning process is carried out in an end-to-end scenario and is not separate. By this our proposed network captures the hidden patterns of seizure types from two different views of the signals, time-frequency wavelet-based features, and raw EEG data.

As we mentioned in the Introduction section, enabling a deep learning model to view the data from different aspects and representations leverages the model's performance and robustness. In the proposed network, the CNN sub-model fed with the wavelet-based features extracts local features of the seizure type, such as transient events and spike activity, while the Bi-LSTM sub-model with its attention mechanism extracts long-term and short-term dependency features from the raw EEG data, allowing the model to understand the overall structure and dynamics of the signal. We also utilized each sub-model as an independent model and added a final softmax classification layer to each for comparison with the proposed MP-SeizNet.

## 3. Results

In this section, we report the findings from our experimental approach to multi-class seizure type classification. We start with the evaluation metrics used with EEG data from the TUSZ dataset, and then reveal the outcomes of each individual sub-model and the final proposed MP-SeizNet. We then compare our results with the existing state of the art solutions.

### 3.1. Training and performance evaluation

Extra care was taken during the evaluation of our proposed technique to ensure its validity. Due to the class-imbalance problem found in the TUSZ dataset, the accuracy metric alone is not going to be helpful. Therefore, we opted for the weighted F1-score to evaluate our proposed approach which is the metric that was used in all existing state-of-the-art researches [6, 41, 48, 42, 51, 28, 44, 52, 45]. The weighted F1-score is presented as [42]:

$$\text{Weighted } F1 = \sum_{i=1}^{5} \frac{2 \times \text{precision}_i \times \text{recall}_i}{\text{precision}_i + \text{recall}_i} \times w_i \quad (11)$$

where $w_i$ is the weight of the $i_{th}$ class depending on the number of positive examples in that class. Moreover, statistical tests were employed to ensure the reliability and effectiveness of the proposed model. The overall sensitivity, specificity, positive predictive value (PPV), negative predictive (NPV), and the area under the curve (AUC), are derived from the confusion matrices [66]. PYCM Python library was utilized to obtain these measures [67]. Additionally, the performances of the proposed models were compared at 95% confidence intervals (CI).

The evaluation criteria are split into two different scenarios seizure-wise and patient-wise cross-validation. In the following two sections, we explain in details the implementation, advantages and disadvantages of each scenario.

#### 3.1.1. Seizure-wise cross-validation

Cross-validation (CV) is a validation technique that uses different subsets of data to test and train a model across several rounds to ensure its effectiveness. The first scenario is the evaluation across seizures without paying attention to patient-specific data, which is inspired by the state-of-the-art technique [41]. In this evaluation technique, the dataset is split into several folds, each having a proportional distribution of each class (label) in the entire dataset. The drawback of this technique is that data from one patient could exist in the training, validation, and test datasets. Despite this drawback, researchers assume that each seizure event holds sufficient data variations to represent the seizure regardless of the the personal similarities [45]. Moreover, the data from the same individual may vary over time, as is the nature of EEG technology [55]. The benefit of this evaluation scheme is that it can assist neurologists in precisely localizing the status of seizure events drift contained in EEG signals and across EEG channels for a specific patient [1].

In our experiments, we used a five-fold cross-validation technique. The entire dataset is randomly and equally split into five folds; each fold holds a proportional distribution of the classes in the entire dataset. The model is trained on four folds and then assessed on the fifth fold. This process is repeated until each fold is utilized as a test set. The training data is further divided into training and validation subsets with 80:20, respectively. At the end, the average F1-score is calculated over all five folds.

#### 3.1.2. Patient-wise cross-validation

The second validation scheme evaluates the proposed technique for different patients, also known as the patient-independent validation technique [6, 28]. This scenario addresses the issue of a model's generalizability to a new group of patients after being trained on data from a previous group of patients. The drawback of this scenario is that a model must learn from the limited number of patient data available, which may limit the data for training deep learning networks as they are being data-hungry. In contrast, the advantages of this scenario is that it ensures the validity of the proposed automated system to be applicable for real-world scenarios as the new data always comes from unseen new patient data.

In our experiments, three folds cross-validation was used as suggested by Roy et al. algorithm [6]. Since the selected seizure classes in this study are recorded from at least three patients, see Table 2. Therefore, the data is split into three folds, where each fold has data from different patients. During training, the model is trained on two folds and tested on the remaining third fold. This process is repeated until all folds are used as a test set. The overall performance is reported using the averaged F-1 score for all folds.

In the TUSZ, there is a huge variance in the number of EEG data for each patient. Some patients have multiple recordings while some only have one. This makes the test data sometimes greater in number than the training data for few classes. In these cases, if the test data for certain seizure class is significantly larger than the training data for the same seizure type, we randomly take a subset of the test data for that class, ensuring to cover all parts of the seizure event from the start, the middle and the end of the event.

For training, all of the networks are trained for 100 epochs with batch-size set to 64. As a way to prevent over-fitting during training, early stopping is used as a method of regularization. This method keeps an eye on the validation loss and stops training if the loss doesn't improve over the course of 10 epochs. The cross-entropy is used as a cost function with Adam as the optimization solver. The initial learning rate is set to 0.01, and with the use of callback, the learning rate is tuned during training and its value is reduced based on the loss function, with minimum value set to $10^{-4}$.

### 3.2. Experimental results

To demonstrate the effectiveness of our proposed novel approach, we compare the results of our proposed networks with sub-models of CNN and Bi-LSTM. Table 5 presents the classification results in terms of weighted average F1-score. It can be clearly observed that the proposed neural network achieves an F1-score of 87.62%, outperforming both the CNN and Bi-LSTM-attention by more than 1 to 3% in patient-wise cross-validation scenario. This shows the importance of both sub-models, each contributing with a set of important features from which the final model is benefiting from them. Moreover, the results of both sub-models show very competitive results compared to each other. These results prove the suitability of both proposed models in solving the problem of patient-wise seizure types classification.

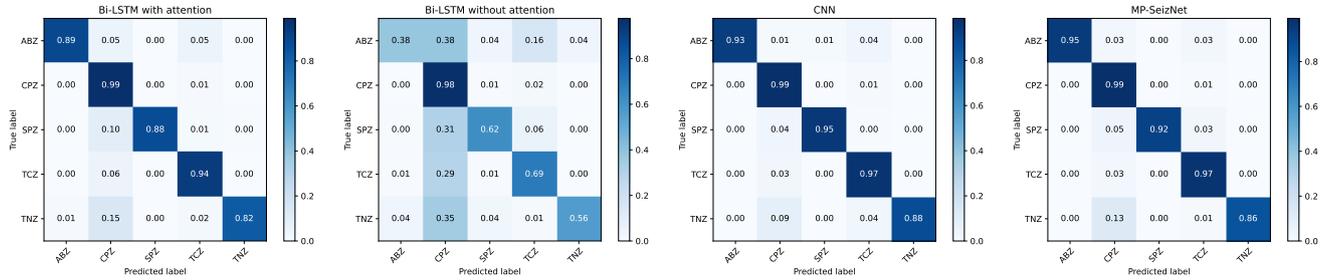

**Fig. 6**: Normalized confusion matrices of specific seizure type classification for sub-model CNN, Bi-LSTM-with attention, Bi-LSTM without attention and the proposed neural network models evaluated on the TUSZ v1.5.2 dataset.

On the other hand, at the seizure level, the results in Table 5 confirm the robustness of our proposed novel approach in classifying seizures type, achieving an F1-score of 98.01%. It is worth noting that the CNN sub-model, which was employed for comparison, produced results that were very similar, with a difference of less than 0.1%. This indicates the efficacy of our approach in capturing essential information from the raw EEG signals with the crucial role played by wavelet-based features. Table 6 presents the highest achievements of the proposed MP-SeizNet and the sub-models CNN and Bi-LSTM for different statistical measures. Additionally, the 95% confidence interval of our proposed MP-SeizNet was computed using 20 randomized tests, achieving (F1-score:0.976, CI:0.975–0.977) and (F1-score: 0.963, CI:0.961–0.965).

Lastly, the attention-based Bi-LSTM sub-model has proven to be an effective tool in the analysis of EEG signals. With an F1-score of 97.04%, it accurately identifies specific seizure types within raw EEG data. The attention mechanism integrated into the sub-model significantly enhances its performance, as shown in Fig. 6. The attention mechanism results in an improvement of more than 10% compared to traditional Bi-LSTM models without this feature. These findings demonstrate the usefulness of the attention-based Bi-LSTM sub-model for the analysis of EEG signals in the field of epileptic seizure classification.

Table 1 summarises the research on automated seizure type classification, and to the best of our knowledge, our study is the first to demonstrate such a high accuracy across different patients. using a multi-path neural network.

## 4. Discussion

Our proposed neural network got a high average F1-scores of 98.1% and 87.62% at the seizure and patient lev-els, respectively. It should be emphasized that CNN-based models, especially with pre-processed signals based on FFT or STFT, were often used for multi-class seizure classification [39, 40, 48, 41]. Our CNN-based model, with wavelet-based features, solely outperforms all previous CNN-based research for seizure-type classification. Also, it can be noted from Table 1 that the second most popular approach is the use of RNN-based models, which were employed for seizure classification in these studies [42, 41, 44, 53]. Our proposed model using the Bi-LSTM with attention mechanism obtained higher results than existing RNN-based state-of-the-art studies. Furthermore, to the best of our knowledge, our study is the first to use Bi-LSTM+attention for multi-class seizure type classification using unprocessed raw EEG data. It is possible to deduce from Table 1 that Baghdadi et al. [44] proposed a channel-wise attention-based LSTM for the same problem, achieving an F1-score of 96.87. In comparison, our attention-based Bi-LSTM sub-model obtained higher results with a significantly less complex network. Unlike Baghdadi et al. [44] and Priyasad et al. [45], employing an attention mechanism for each channel independently, we used the attention mechanism in a way that can find the most prevalent features, including the distribution of events from all channels [65]. Fig. 6 compares our Bi-LSTM with and without attention, and it is clear that the attention mechanism helped the model to capture small details that discriminate between seizure events in EEG, especially for those seizure types which have very small training data (ABZ and SPZ).

**Table 5**
Classification results in terms of average F1-scores for sub-model CNN, Bi-LSTM with attention and the proposed neural network.

| Model | Patient Wise Ave. F1-score | Seizure Wise Ave. F1-score |
| --- | --- | --- |
| CNN sub-model | 86.1 | 98.1 |
| Bi-LSTM sub-model | 84.63 | 97.04 |
| MP-SeizNet | 87.62 | 98.01 |

Regarding patient-wise cross-validation, only a few studies considered this assessment aspect. Table 7 compares the results of our proposed approach with state-of-the-art published research. Our novel technique surpasses all existing

**Table 6**
The classification performances (%) for CNN, Bi-LSTM with attention and the proposed MP-SeizNet.

| Model | Sensitivity (CI) | Specificity (CI) | Cohen's Kappa (CI) | PPV (CI) | NPV (CI) | AUC (CI) |
|---|---|---|---|---|---|---|
| CNN sub-model | 98.2 (97.9–98.4) | 96.6 (96.4–96.9) | 94.9 (94.3–95.5) | 98.2 (97.9–98.4) | 96.9 (96.2–97.5) | 97.4 (97.2–97.6) |
| Bi-LSTM sub-model | 97.07 (96.8–98.3) | 93.3 (92.6–94.1) | 91.6 (90.9–92.3) | 97.1 (96.8–97.2) | 96.4 (95.8–96.9) | 95.2 (94.7–95.6) |
| MP-SeizNet | 98.02 (97.8–98.2) | 95.6 (94.9–96.3) | 94.4 (93.7–95) | 98.01 (97.8–98.2) | 97.6 (97.4–97.8) | 96.8 (96.4–97.3) |

CI: 95% Confidence Intervals (Lower bound–Upper bound), PPV: Positive Predictive Value, NPV: Negative Predictive Value

**Table 7**
Patient-wise cross-validation performance comparison with studies found in the literature.

| Study | year | Classes | Patient_Wise CV(%) |
|---|---|---|---|
| Roy et al., [6] | 2020 | 7 | 54.2 F1-score |
| Asif et al., [48] | 2020 | 7 | 62.0 F1-score |
| Albaqami et al., [26] | 2021 | 7 | 64.0 F1-score |
| Albaqami et al., [26] | 2021 | 5 | 66.6 F1-score |
| Albaqami et al., [28] | 2022 | 5 | 74.7 F1-score |
| Jia et al. [50] | 2022 | 7 | 54.2 Accuracy |
| Tang et al., [52] | 2022 | 4 | 74.9 F1-score |
| | | | |
| CNN sub-model | | 5 | 86.1 F1-score |
| Bi-LSTM sub-model | | 5 | 84.6 F1-score |
| MP-SeizNet | | 5 | 87.6 F1-score |

published state-of-the-art studies by a significant margin of 12%, achieving an average F1-score of 87.6%. Additionally, when evaluating both sub-models independently, they either performed better than the models presented in the existing literature and showed very competitive results as compared to our proposed neural network. This result is achieved through two important aspects; first, the wavelet-based features are responsible for capturing local features of the signal, such as the transient events, including spikes and sharp waves[24]. Moreover, the CNN-based sub-model at the higher layers can capture high-level spatial features of the data. Second, our proposed Bi-LSTM sub-model is basically detecting the context of the signal by looking at the raw EEG data, as it is good at finding the sensitive features of the signal. Therefore, our proposed model can capture both local and global sensitive features of the signal, allowing it to generalize better to data from different patients.

As it was suggested in [42, 26, 28, 52] that the unspecific labels found in the TUSZ are not medically disjoint from other specific labels, we ignored the unspecific classes for more realistic situations for specific seizure classification. Unlike [33], where the author tried to re-label the signals into combined labels, we ignored re-labeling of the signals because this must be done with a qualified neurologist. Therefore, in this research, as discussed in the introduction, our only focus is the classification of specific seizure types classes labeled and reviewed by neurologists at Temple Hospital University.

Although the result of our proposed approach in this study at the seizure level is not better than what has been reported in [28], our proposed technique achieves very high results at the patient level compared to the same study. We can relate this to the fact that the amount of data used for training in [28] is more than what we have used in our study, as we split the training data in each fold into 80/20 training and validation (see Section 3.1.1). Shankar et al. [53] reported an F1 of 98.8%, slightly higher than our obtained results. However, their validation criteria is based on a random holdout test set extracted from the whole dataset. They added a seizure-free class to the dataset, which is easier to detect by the model as it is very different from those signals with seizures. Also, their method for feature extraction from EEG data is computationally expensive compared to our proposed technique. Moreover, the 50% overlapping technique in their proposed model might affect the obtained results [28].

There are some limitations of our work. Firstly the training of neural networks is time consuming process but needed to be done once. Secondly, there are couple of classes in the TUSZ dataset which have a small number of patients. In the case of more patients, the proposed neural network is expected to work better. However, there are advantages of our proposed model which are as follows:

- The performance of our model is robust when generalized on unseen patient data using limited training data.

- Our proposed model can be used for other classification tasks, for example, abnormality detection using EEG data.

- Our model can capture local and temporal seizure patterns using wavelet-based features and Bi-LSTM.

## 5. Conclusion

Epilepsy, characterized by sudden seizures, is highly associated with significant morbidity and mortality. These seizures come in different types, and their accurate identification is one of the first steps in the treatment journey for epileptic patients. Given the difficulties associated with manual seizure diagnosis, developing a reliable and automatic seizure-type classification system is vital. This paper proposes a novel deep neural network MP-SeizNet for classifying seizure types: complex partial, simple partial, absence, tonic, and tonic-clonic. Our proposed MP-SeizNet consists of two sub-models: CNN and Bi-LSTM. For the CNN sub-model, features are extracted from the signal based on wavelet transform, and provided as an input. Whereas, the raw EEG signals are used to train the attention-based Bi-LSTM sub-model. The output of the two sub-models is finally merged for classification.

We evaluated our proposed model on the most extensive available EEG database, the TUSZ v1.5.2. We achieved the average F1 scores of 98.1% and 87.6% for patient-specific and patient-non-specific data, respectively. Furthermore, the robustness of our model is validated using five-fold cross-validation. To the best of our knowledge, this is the first study to have generalized the learning process across different subjects with this high accuracy. In the future, we plan to test different variants of deep learning methods with different representations of the EEG data for more accurate and reliable results for small seizure classes.

## Consent for publication

All authors have read and approved the manuscript.

## Declaration of competing interest

The authors have no conflicts of interest to declare.

## Credit authorship contribution statement

**Hezam Albaqami:** Investigation, Data Curation, Visualization, Software, Methodology, Writing - Original Draft, Writing - Review & Editing, Validation. **Ghulam Mubashar Hassan:** Conceptualization, Writing - Review & Editing, Supervision. **Amitava Datta:** Conceptualization, Writing - Review & Editing, Project administration, Resources, Supervision.

## Acknowledgments

This work was supported by a scholarship from University of Jeddah, 201589, Saudi Arabia.

## References

[1] D. P. Yedurkar, S. P. Metkar, T. Stephan, Multiresolution directed transfer function approach for segment-wise seizure classification of epileptic eeg signal, Cognitive Neurodynamics (2022) 1–15.

[2] R. S. Fisher, W. V. E. Boas, W. Blume, C. Elger, P. Genton, P. Lee, J. Engel Jr, Epileptic seizures and epilepsy: definitions proposed by the international league against epilepsy (ilae) and the international bureau for epilepsy (ibe), Epilepsia 46 (2005) 470–472.

[3] A. Subasi, J. Kevric, M. Abdullah Canbaz, Epileptic seizure detection using hybrid machine learning methods, Neural Computing and Applications 31 (2019) 317–325. doi:10.1007/s00521-017-3003-y.

[4] N. Alessi, P. Perucca, A. M. McIntosh, Missed, mistaken, stalled: Identifying components of delay to diagnosis in epilepsy, Epilepsia 62 (2021) 1494–1504.

[5] M. M. Zack, R. Kobau, National and state estimates of the numbers of adults and children with active epilepsy—united states, 2015, MMWR Morb Mortal Wkly Rep 66 (2017) 821.

[6] S. Roy, U. Asif, J. Tang, S. Harrer, Seizure type classification using EEG signals and machine learning: Setting a benchmark, in: Proceedings of the IEEE Signal Processing in Medicine and Biology Symposium, 2020, pp. 1–6.

[7] A. Mobed, M. Shirafkan, S. Charsouei, J. Sadeghzadeh, A. Ahmadalipour, Biosensors technology for anti-epileptic drugs, Clinica Chimica Acta 533 (2022) 175–182. URL: https://www.sciencedirect.com/science/article/pii/S0009898122012177. doi:https://doi.org/10.1016/j.cca.2022.06.027.

[8] M. J. Casale, L. V. Marcuse, J. J. Young, N. Jette, F. E. Panov, H. A. Bender, A. E. Saad, R. S. Ghotra, S. Ghatan, A. Singh, et al., The sensitivity of scalp eeg at detecting seizures—a simultaneous scalp and stereo eeg study, Journal of Clinical Neurophysiology 39 (2022) 78–84.

[9] S. Sanei, J. A. Chambers, EEG signal processing, John Wiley & Sons, 2013.

[10] I. E. Scheffer, S. Berkovic, G. Capovilla, M. B. Connolly, J. French, L. Guilhoto, E. Hirsch, S. Jain, G. W. Mathern, S. L. Moshé, et al., ILAE classification of the epilepsies: position paper of the ILAE Commission for Classification and Terminology, Epilepsia 58 (2017) 512–521.

[11] I. Obeid, J. Picone, Machine learning approaches to automatic interpretation of EEGs, Signal processing and machine learning for biomedical big data (2018) 70. doi:https://www.isip.piconepress.com/publications/book_sections/2017/crc_press/auto_eeg/.

[12] A. Burton, How do we fix the shortage of neurologists?, The Lancet Neurology 17 (2018) 502–503.

[13] L. Parviainen, R. Kälviäinen, L. Jutila, Impact of diagnostic delay on seizure outcome in newly diagnosed focal epilepsy, Epilepsia open 5 (2020) 605–610.

[14] C. Panayiotopoulos, Optimal use of the EEG in the diagnosis and management of epilepsies, in: The epilepsies: seizures, syndromes and management, Bladon Medical Publishing, 2005.

[15] T. U. Syed, W. C. LaFrance Jr, E. S. Kahriman, S. N. Hasan, V. Rajasekaran, D. Gulati, S. Borad, A. Shahid, G. Fernandez-Baca, N. Garcia, et al., Can semiology predict psychogenic nonepileptic seizures? a prospective study, Annals of neurology 69 (2011) 997–1004.

[16] L. A. Gemein, R. T. Schirrmeister, P. Chrabąszcz, D. Wilson, J. Boedecker, A. Schulze-Bonhage, F. Hutter, T. Ball, Machine-learning-based diagnostics of eeg pathology, NeuroImage 220 (2020) 117021.

[17] J. Kevric, A. Subasi, Comparison of signal decomposition methods in classification of EEG signals for motor-imagery BCI system, Biomedical Signal Processing and Control 31 (2017) 398–406.

[18] G. Altan, G. Inat, Eeg based spatial attention shifts detection using time-frequency features on empirical wavelet transform, Akıllı Sistemler ve Uygulamaları Dergisi (Journal of Intelligent Systems with Applications) 4 (2021) 144–149. doi:10.54856/jiswa.202112181.

[19] T. N. Alotaiby, S. A. Alshebeili, F. M. Alotaibi, S. R. Alrshoud, Epileptic seizure prediction using CSP and LDA for scalp EEG signals, Computational intelligence and neuroscience 2017 (2017).

[20] S. Altunay, Z. Telatar, O. Erogul, Epileptic EEG detection using the linear prediction error energy, Expert Systems with Applications 37 (2010) 5661–5665.

[21] F. O.K., R. R., Time-domain exponential energy for epileptic eeg signal classification, Neuroscience Letters 694 (2019) 1–8. URL: https://www.sciencedirect.com/science/article/pii/S0304394018307444. doi:https://doi.org/10.1016/j.neulet.2018.10.062.

[22] K. Polat, S. Güneş, Classification of epileptiform EEG using a hybrid system based on decision tree classifier and fast fourier transform, Applied Mathematics and Computation 187 (2007) 1017–1026.

[23] A. Subasi, EEG signal classification using wavelet feature extraction and a mixture of expert model, Expert Systems with Applications 32 (2007) 1084–1093.

[24] H. Albaqami, G. M. Hassan, A. Subasi, A. Datta, Automatic detection of abnormal EEG signals using wavelet feature extraction and gradient boosting decision tree, Biomedical Signal Processing and Control 70 (2021) 102957.

[25] R. Sarić, D. Jokić, N. Beganović, L. G. Pokvić, A. Badnjević, Fpga-based real-time epileptic seizure classification using artificial neural network, Biomedical Signal Processing and Control 62 (2020) 102106.

[26] H. Albaqami, G. Hassan, A. Datta, Comparison of wpd, dwt and dtcwt for multi-class seizure type classification, in: 2021 IEEE Signal Processing in Medicine and Biology Symposium (SPMB), IEEE, 2021, pp. 1–7.

[27] N. McCallan, S. Davidson, K. Y. Ng, P. Biglarbeigi, D. Finlay, B. L. Lan, J. McLaughlin, Seizure classification of eeg based on wavelet

signal denoising using a novel channel selection algorithm, in: 2021 Asia-Pacific Signal and Information Processing Association Annual Summit and Conference (APSIPA ASC), IEEE, 2021, pp. 1269–1276.

[28] H. Albaqami, G. M. Hassan, A. Datta, Wavelet-based multi-class seizure type classification system, Applied Sciences 12 (2022) 5702.

[29] A. Subasi, S. Jukic, J. Kevric, Comparison of EMD, DWT and WPD for the localization of epileptogenic foci using random forest classifier, Measurement 146 (2019) 846–855.

[30] R. T. Schirrmeister, J. T. Springenberg, L. D. J. Fiederer, M. Glasstetter, K. Eggensperger, M. Tangermann, F. Hutter, W. Burgard, T. Ball, Deep learning with convolutional neural networks for eeg decoding and visualization, Human brain mapping 38 (2017) 5391–5420.

[31] G. Altan, A. Yayık, Y. Kutlu, Deep learning with convnet predicts imagery tasks through eeg, Neural Processing Letters 53 (2021) 2917–2932.

[32] S. Roy, I. Kiral-Kornek, S. Harrer, Chrononet: a deep recurrent neural network for abnormal eeg identification, in: Artificial Intelligence in Medicine: 17th Conference on Artificial Intelligence in Medicine, AIME 2019, Poznan, Poland, June 26–29, 2019, Proceedings 17, Springer, 2019, pp. 47–56.

[33] R. Schirrmeister, L. Gemein, K. Eggensperger, F. Hutter, T. Ball, Deep learning with convolutional neural networks for decoding and visualization of eeg pathology, in: 2017 IEEE Signal Processing in Medicine and Biology Symposium (SPMB), 2017, pp. 1–7. doi:10.1109/SPMB.2017.8257015.

[34] G. Altan, Y. Kutlu, Generative autoencoder kernels on deep learning for brain activity analysis, Natural and Engineering Sciences 3 (2018) 311–322. doi:10.28978/nesciences.468978.

[35] X. Tian, Z. Deng, W. Ying, K.-S. Choi, D. Wu, B. Qin, J. Wang, H. Shen, S. Wang, Deep multi-view feature learning for eeg-based epileptic seizure detection, IEEE Transactions on Neural Systems and Rehabilitation Engineering 27 (2019) 1962–1972. doi:10.1109/TNSRE.2019.2940485.

[36] Y. Yuan, G. Xun, K. Jia, A. Zhang, A multi-view deep learning method for epileptic seizure detection using short-time fourier transform, in: Proceedings of the 8th ACM International Conference on Bioinformatics, Computational Biology, and Health Informatics, ACM-BCB '17, Association for Computing Machinery, New York, NY, USA, 2017, p. 213–222. URL: https://doi.org/10.1145/3107411.3107419. doi:10.1145/3107411.3107419.

[37] Y. Li, M. Yang, Z. Zhang, A survey of multi-view representation learning, IEEE transactions on knowledge and data engineering 31 (2018) 1863–1883.

[38] Y. Jiang, Y. Zhang, C. Lin, D. Wu, C.-T. Lin, Eeg-based driver drowsiness estimation using an online multi-view and transfer tsk fuzzy system, IEEE Transactions on Intelligent Transportation Systems 22 (2021) 1752–1764. doi:10.1109/TITS.2020.2973673.

[39] N. Sriraam, Y. Temel, S. V. Rao, P. L. Kubben, et al., A convolutional neural network based framework for classification of seizure types, in: 2019 41st Annual International Conference of the IEEE Engineering in Medicine and Biology Society (EMBC), IEEE, 2019, pp. 2547–2550.

[40] S. Raghu, N. Sriraam, Y. Temel, S. V. Rao, P. L. Kubben, EEG based multi-class seizure type classification using convolutional neural network and transfer learning, Neural Networks 124 (2020) 202–212.

[41] T. Liu, N. D. Truong, A. Nikpour, L. Zhou, O. Kavehei, Epileptic seizure classification with symmetric and hybrid bilinear models, IEEE Journal of Biomedical and Health Informatics 24 (2020) 2844–2851. doi:10.1109/JBHI.2020.2984128. arXiv:2001.06282.

[42] D. Ahmedt-Aristizabal, T. Fernando, S. Denman, L. Petersson, M. J. Aburn, C. Fookes, Neural memory networks for seizure type classification, in: 42nd Annual International Conference of the IEEE Engineering in Medicine & Biology Society (EMBC), IEEE, 2020.

[43] A. Shankar, S. Dandapat, S. Barma, Seizure type classification using eeg based on gramian angular field transformation and deep learning, in: 2021 43rd Annual International Conference of the IEEE Engineering in Medicine & Biology Society (EMBC), 2021, pp. 3340–3343. doi:10.1109/EMBC46164.2021.9629791.

[44] A. Baghdadi, S. Daoud, M. Dammak, C. Mhiri, P. Siarry, A. M. Alimi, et al., A channel-wise attention-based representation learning method for epileptic seizure detection and type classification (2021).

[45] D. Priyasad, T. Fernando, S. Denman, S. Sridharan, C. Fookes, Interpretable seizure classification using unprocessed eeg with multi-channel attentive feature fusion, IEEE Sensors Journal 21 (2021) 19186–19197.

[46] I. R. D. Saputro, N. D. Maryati, S. R. Solihati, I. Wijayanto, S. Hadiyoso, R. Patmasari, Seizure type classification on EEG signal using support vector machine, in: Journal of Physics: Conference Series, volume 1201, IOP Publishing, 2019, p. 012065.

[47] I. Wijayanto, R. Hartanto, H. A. Nugroho, B. Winduratna, Seizure type detection in epileptic EEG signal using empirical mode decomposition and support vector machine, in: Proceedings - 2019 International Seminar on Intelligent Technology and Its Application, ISITIA 2019, Institute of Electrical and Electronics Engineers Inc., 2019, pp. 314–319. doi:10.1109/ISITIA.2019.8937205.

[48] U. Asif, S. Roy, J. Tang, S. Harrer, SeizureNet: Multi-spectral deep feature learning for seizure type classification, in: Machine Learning in Clinical Neuroimaging and Radiogenomics in Neuro-oncology, Springer, 2020, pp. 77–87.

[49] S. Naze, J. Tang, J. R. Kozloski, S. Harrer, Features importance in seizure classification using scalp eeg reduced to single timeseries, in: 2021 43rd Annual International Conference of the IEEE Engineering in Medicine & Biology Society (EMBC), 2021, pp. 329–332. doi:10.1109/EMBC46164.2021.9630398.

[50] G. Jia, H.-K. Lam, K. Althoefer, Variable weight algorithm for convolutional neural networks and its applications to classification of seizure phases and types, Pattern Recognition 121 (2022) 108226. URL: https://www.sciencedirect.com/science/article/pii/S0031320321004076. doi:https://doi.org/10.1016/j.patcog.2021.108226.

[51] S. Zhang, G. Liu, R. Xiao, W. Cui, J. Cai, X. Hu, Y. Sun, J. Qiu, Y. Qi, A combination of statistical parameters for epileptic seizure detection and classification using vmd and nltwsvm, Biocybernetics and Biomedical Engineering 42 (2022) 258–272.

[52] S. Tang, J. Dunnmon, K. K. Saab, X. Zhang, Q. Huang, F. Dubost, D. Rubin, C. Lee-Messer, Self-supervised graph neural networks for improved electroencephalographic seizure analysis, in: International Conference on Learning Representations, 2022. URL: https://openreview.net/forum?id=k9bx1EfHI_-.

[53] A. Shankar, S. Dandapat, S. Barma, Seizure types classification by generating input images with in-depth features from decomposed eeg signals for deep learning pipeline, IEEE Journal of Biomedical and Health Informatics (2022) 1–1. doi:10.1109/JBHI.2022.3159531.

[54] R. S. Fisher, J. H. Cross, J. A. French, N. Higurashi, E. Hirsch, F. E. Jansen, L. Lagae, S. L. Moshé, J. Peltola, E. Roulet Perez, I. E. Scheffer, S. M. Zuberi, Operational classification of seizure types by the international league against epilepsy: Position paper of the ILAE commission for classification and terminology, Epilepsia 58 (2017) 522–530. doi:10.1111/epi.13670.

[55] I. Obeid, J. Picone, The temple university hospital EEG data corpus, Frontiers in neuroscience 10 (2016) 196.

[56] V. Shah, E. Von Weltin, S. Lopez, J. R. McHugh, L. Veloso, M. Golmohammadi, I. Obeid, J. Picone, The temple university hospital seizure detection corpus, Frontiers in neuroinformatics 12 (2018) 83.

[57] H. H. Jasper, The ten-twenty electrode system of the international federation, Electroencephalogr. Clin. Neurophysiol. 10 (1958) 370–375.

[58] N. Saito, Frequently asked questions on wavelets (2004).

[59] G. Chen, Automatic eeg seizure detection using dual-tree complex wavelet-fourier features, Expert Systems with Applications 41 (2014) 2391–2394. doi:https://doi.org/10.1016/j.eswa.2013.09.037.

[60] T. H. Cormen, C. E. Leiserson, R. L. Rivest, C. Stein, Introduction to algorithms, MIT press, 2009.

[61] A. Shoeibi, N. Ghassemi, M. Khodatars, P. Moridian, R. Alizadehsani, A. Zare, A. Khosravi, A. Subasi, U. R. Acharya, J. M. Gorriz, Detection of epileptic seizures on eeg signals using anfis classifier, autoencoders and fuzzy entropies, Biomedical Signal Processing and Control 73 (2022) 103417.

[62] A. Shoeibi, N. Ghassemi, R. Alizadehsani, M. Rouhani, H. Hosseini-Nejad, A. Khosravi, M. Panahiazar, S. Nahavandi, A comprehensive comparison of handcrafted features and convolutional autoencoders for epileptic seizures detection in EEG signals, Expert Systems with Applications 163 (2021) 113788. doi:https://doi.org/10.1016/j.eswa.2020.113788.

[63] A. Géron, Hands-on machine learning with Scikit-Learn, Keras, and TensorFlow: Concepts, tools, and techniques to build intelligent systems, " O'Reilly Media, Inc.", 2019.

[64] Y. Hou, L. Zhou, S. Jia, X. Lun, A novel approach of decoding eeg four-class motor imagery tasks via scout esi and cnn, Journal of neural engineering 17 (2020) 016048.

[65] G. Zhang, V. Davoodnia, A. Sepas-Moghaddam, Y. Zhang, A. Etemad, Classification of hand movements from eeg using a deep attention-based lstm network, IEEE Sensors Journal 20 (2020) 3113–3122. doi:10.1109/JSEN.2019.2956998.

[66] G. Altan, Deepoct: An explainable deep learning architecture to analyze macular edema on oct images, Engineering Science and Technology, an International Journal 34 (2022) 101091. doi:https://doi.org/10.1016/j.jestch.2021.101091.

[67] S. Haghighi, M. Jasemi, S. Hessabi, A. Zolanvari, PyCM: Multiclass confusion matrix library in python, Journal of Open Source Software 3 (2018) 729. doi:10.21105/joss.00729.